\documentstyle[aps,preprint]{revtex}
\begin{document}
\draft{}
\title{ ON THE NATURE OF THE SCALAR $a_0(980)$ AND $f_0(980)$-MESONS
\thanks{Talk given at INTERNATIONAL WORKSHOP $e^+e^-$ COLLISIONS
FROM $\phi$ TO $J/\psi$,
Budker Institute of Nuclear Physics,
Novosibirsk, Russia, March 1-5, 1999}}
\author{N.N. Achasov \\
Laboratory of Theoretical Physics, \\
S.L. Sobolev Institute for Mathematics,\\
630090 Novosibirsk 90, Russia\\[1pc]
E-mail: achasov@math.nsc.ru}
\date{\today}
\maketitle
\begin{abstract}
It is presented a critical consideration of all unusual properties of the
scalar $a_0(980)$ and $f_0(980)$-mesons in the four-quark, two-quark and
molecular models. The arguments are adduced that the four-quark model is more
preferable. It is discussed the complex of experiments that could finally
resolve this issue.
\end{abstract}

\tableofcontents
\section{Introduction}
Spherical Neutral Detector (SND) from the $e^+e^-$-collider VEPP-2M in
Novosibirsk has discovered \cite{snda098,sndf098} the electric dipole decays
$\phi\to\gamma\pi^0\pi^0$ and $\phi\to\gamma\pi^0\eta$ in the region of the
soft by strong interaction standard photons with the energy
$\omega < 120\,\mbox{MeV}$, i.e., in the region of the scalar $a_0(980)$ and
$f_0(980)$-mesons  $m_{\pi^0\pi^0}> 900\,
\mbox{MeV}$ and $m_{\pi^0\eta}> 900\,\mbox{MeV}$,
$\phi\to\gamma f_0(980)\to\gamma\pi^0\pi^0$ and
$\phi\to\gamma a_0(980)\to\gamma\pi^0\eta$. The data are
\begin{eqnarray}
\label{snd1}
&& B(\phi\to\gamma\pi^0\pi^0\,;\,m_{\pi^0\pi^0}>900\,\mbox{MeV})=
(0.5\pm 0.06\pm 0.06)\cdot 10^{-4}\nonumber\\
&&\mbox{at total}\ \ B(\phi\to\gamma\pi^0\pi^0)=
(1.14\pm 0.10\pm 0.12)\cdot 10^{-4}\,,\\
\label{snd2}
&& B(\phi\to\gamma\pi^0\eta\,;\,m_{\pi^0\eta}>900\,\mbox{MeV})\simeq
0.5\cdot 10^{-4}\,,\nonumber\\
&&\mbox{at total}\ \ B(\phi\to\gamma\pi^0\eta)=(0.83\pm 023)\cdot 10^{-4}\,.
\end{eqnarray}

Cryogenic Magnetic Detector-2 (CMD-2) from the $e^+e^-$-collider
VEPP-2M in Novosibirsk has confirmed these results \cite{solodov-99}.

New preliminary SND analysis \cite{golubev-99} of 1998 data with double
statistics in comparison to 1996 data \cite{snda098} also confirms the
results (\ref{snd2}).

The branching ratios in Eqs. (\ref{snd1}) and (\ref{snd2}) are great for this
photon energy region and, probably,  can be understood only if four-quark
resonances are produced \cite{achasov-89,achasov-97}.
Note that the $a_0(980)$ meson is produced in the $\phi$ radiative
decay as intensively as the containing strange quarks $\eta'$ meson.

\section{Evidences for strange quarks in the \lowercase{$f_0(980)$} and
\lowercase{$a_0(980)$}}

To feel why numbers in Eqs. (\ref{snd1}) and (\ref{snd2}) are great, one can
adduce the rough estimate. Let there be structural radiation without a
resonance in the final state with the spectrum
$$\frac{d\Gamma(\phi\to\gamma\pi^0\pi^0(\eta))}{d\omega}\sim
\frac{\alpha}{\pi}\cdot\delta_{OZI}\cdot\frac{1}{m^3_{\phi}}\omega^3\,,$$
where $\delta_{OZI}\sim 10^{-2}$ is a factor describing the suppression
by Okubo-Zweig-Iizuki (OZI) rule.
Recall that the $\omega^3$ law follows from gauge invariance.
Really, the decay amplitude is proportional to the electromagnetic field
$F_{\mu\nu}$ ( in our case to the electric field ), i. e., to the photon
energy $\omega$ in the soft photon region.

Then one gets
\begin{eqnarray}
&&\Gamma(\phi\to\gamma\pi^0\pi^0(\eta))\sim\frac{1}{4}\cdot\frac{\alpha}{\pi}
\cdot\delta_{OZI}\cdot\frac{\omega^4_0}{m^3_{\phi}}\simeq 10^{-6}\,
\mbox{MeV}\,,\nonumber\\
&&\mbox{i.e.}\ \ B(\phi\to\gamma\pi^0\pi^0(\eta))\sim 2\cdot 10^{-7}\,,
\nonumber
\end{eqnarray}
where $\omega_0=120\,\mbox{MeV}$.

To understand, why Eq. (\ref{snd2}) points to four-quark model, is particular
easy. Really, the $\phi$-meson is the isoscalar practically pure
$s\bar s$-state, that decays to the isovector hadron state $\pi^0\eta$ and
the isovector photon. The isovector photon originates from the $\rho$-meson,
$\phi\to\rho a_0(980)\to\gamma\pi^0\eta$, the structure of which in this
energy region is familiar
\begin{eqnarray}
\label{rho}
&&\rho\approx (u\bar u-d\bar d)/\sqrt{2}\,.
\end{eqnarray}

The structure of the $a_0(980)$-meson , from which the $\pi^0\eta$-system
originates, in general, is
\begin{eqnarray}
\label{a0}
&& X=a_0(980)=c_1(u\bar u-d\bar d)/\sqrt{2}+
c_2s\bar s(u\bar u-d\bar d)/\sqrt{2} + ...\ .
\end{eqnarray}

The strange quarks, with the first term in Eq. (\ref{a0}) taken as
dominant, are absent in the intermediate state.  So, we would have the
suppressed by OZI-rule decay with
$B(\phi\to\gamma a_0(980)\to\gamma\pi^0\eta)\sim 10^{-6}$ owing to
the real part of the decay amplitude \cite{achasov-97}. The imaginary part
of the decay amplitude, resulted from the $K^+K^-$- intermediate state
($\phi\to\gamma K^+K^-\to\gamma a_0(980)\to\gamma\pi^0\eta$), violates the
OZI-rule and increases the branching ratio \cite{achasov-89,achasov-97} up
to $10^{-5}$.

The four-quark hypothesis is supported also by the $J/\psi$-decays. Really,
\cite{pdg-98}
\begin{eqnarray}
\label{a2rho}
&& B(J/\psi\to a_2(1320)\rho)= (109\pm 22)\cdot 10^{-4}\,,
\end{eqnarray}
while \cite{kopke-89}
\begin{eqnarray}
\label{a0rho}
&& B(J/\psi\to a_0(980)\rho)< 4.4\cdot 10^{-4}\,.
\end{eqnarray}

The suppression
\begin{eqnarray}
\label{a0rho/a2rho}
&& B(J/\psi\to a_0(980)\rho) /B(J/\psi\to a_2(1320)\rho)< 0.04\pm 0.008
\end{eqnarray}
seems strange, if one considers the $a_2(1320)$ and $a_0(980)$-states
as the tensor and scalar two-quark states from the same P-wave multiplet
with the quark structure
\begin{eqnarray}
\label{a0qq}
&& a_0^0=(u\bar u-d\bar d)/\sqrt{2}\ \ ,\ \ a^+_0=u\bar d\ \ ,\  a^-_0
= d\bar u\,.
\end{eqnarray}
While the four-quark nature of the $a_0(980)$-meson with the symbolic
quark structure
\begin{eqnarray}
\label{a0qqqq}
&& a^0_0=s\bar s(u\bar u-d\bar d)/\sqrt{2}\ \ ,\ \ a_0^+=s\bar su\bar d\ \ ,
\ \ a_0^-=s\bar sd\bar u
\end{eqnarray}
is not c®ntrary to the suppression in Eq. (\ref{a0rho/a2rho}).

Besides, it was predicted in \cite{achasov-82} that the production vigor
of the $a_0(980)$-meson, with it taken as the four-quark state from the
lightest nonet of the MIT-bag \cite{jaffe-77}, in the
$\gamma\gamma$-collisions should be suppressed by the value order in
comparison with the $a_0(980)$-meson taken as the two-quark P-wave state.
In the four-quark model there was obtained the estimate  \cite{achasov-82}
\begin{eqnarray}
\label{ga0gg4q}
&& \Gamma(a_0(980)\to\gamma\gamma)\sim 0.27\,\mbox{keV,}
\end{eqnarray}
which was confirmed by experiment \cite{crystalball,jade}
\begin{eqnarray}
\label{ga0ggexp}
&&\Gamma (a_0\to\gamma\gamma)=(0.19\pm 0.07 ^{+0.1}_{-0.07})/B(a_0\to\pi\eta)
\,
\mbox{keV, Crystal Ball,}\nonumber\\
 && \Gamma (a_0\to\gamma\gamma)=(0.28\pm 0.04\pm 0.1)/B(a_0\to\pi\eta)\,
\mbox{keV, JADE.}
\end{eqnarray}
At the same time in the two-quark model (\ref{a0qq}) it was anticipated
\cite{budnev-79,barnes-85} that
\begin{eqnarray}
\label{ga0gg2q}
\Gamma(a_0\to\gamma\gamma)=(1.5 - 5.9)\Gamma (a_2\to\gamma\gamma)=
(1.5 - 5.9)(1.04\pm 0.09)\,\mbox{keV.}
\end{eqnarray}
The wide scatter of the predictions is connected with different reasonable
guesses of the potential form.

As for the $\phi\to \gamma f_0(980)\to\gamma\pi^0\pi^0$-decay, the more
sophisticated analysis is required.

The structure of the $f_0(980)$-meson, from which the $\pi^0\pi^0$-system
originates, in general, is
\begin{eqnarray}
\label{f0}
&& Y=f_0(980)=\tilde c_0gg+\tilde c_1(u\bar u+d\bar d)/\sqrt{2}+
\tilde c_2s\bar s + \tilde c_3s\bar s(u\bar u+d\bar d)/\sqrt{2} + ...\ .
\end{eqnarray}

First we discuss  a possibility to treat the $f_0(980)$-meson as the
quark-antiquark state.

The hypothesis that the $f_0(980)$-meson is the lowest two-quark P-wave
scalar state with the quark structure
\begin{eqnarray}
\label{f0qq}
&& f_0=(u\bar u+d\bar d)/\sqrt{2}
\end{eqnarray}
contradicts Eq. (\ref{snd1}) in view of OZI, much as Eq. (\ref{a0qq})
contradicts Eq. (\ref{snd2}) (see  the above arguments).

Besides, this hypothesis contradicts a variety of facts:\\
i) the strong coupling with the $K\bar K$-channel
\cite{achasov-84,achasov-97}
\begin{eqnarray}
\label{r}
1<R=|g_{f_0K^+K^-}/g_{f_0\pi^+\pi^-}|^2\lesssim 8\,,
\end{eqnarray}
for from Eq. (\ref{f0qq}) it follows that
$|g_{f_0K^+K^-}/g_{f_0\pi^+\pi^-}|^2=\lambda/4\simeq 1/8$, where $\lambda$
takes into account the strange sea suppression;\\
ii) the weak coupling with gluons \cite{eigen-88}
\begin{eqnarray}
\label{f0gluons}
&& B(J/\psi\to\gamma f_0(980)\to\gamma\pi\pi) < 1.4\cdot 10^{-5}
\end{eqnarray}
opposite the expected one \cite{farrar-94} for Eq. (\ref{f0qq})
\begin{eqnarray}
\label{farrar2q}
&& B(J/\psi\to\gamma f_0(980))\gtrsim B(J/\psi\to\gamma f_2(1270))/4\simeq
3.4\cdot 10^{-4}\,;
\end{eqnarray}
iii) the weak coupling with photons \cite{marsiske-90,gidal-88}
\begin{eqnarray}
\label{gf0ggexp}
&&\Gamma (f_0\to\gamma\gamma)=(0.31\pm 0.14\pm 0.09)\,
\mbox{keV, Crystal Ball,}\nonumber\\
&& \Gamma (f_0\to\gamma\gamma)=(0.24\pm 0.06\pm 0.15)\, \mbox{keV, MARK II}
\end{eqnarray}
opposite the expected one \cite{budnev-79,barnes-85} for Eq. (\ref{f0qq})
\begin{eqnarray}
\label{gf0gg2q}
&&\Gamma(f_0\to\gamma\gamma)=(1.7 - 5.5)\Gamma (f_2\to\gamma\gamma)=
(1.7 - 5.5)(2.8\pm 0.4)\,\mbox{keV;}
\end{eqnarray}
iv) the decays $J/\psi\to f_0(980)\omega$, $J/\psi\to f_0(980)\phi$,
$J/\psi\to f_2(1270)\omega$, $J/\psi\to f_2'(1525)\phi$ \cite{pdg-98}
\begin{eqnarray}
\label{f0omega}
B(J/\psi\to f_0(980)\omega)=(1.4\pm 0.5)\cdot 10^{-4}\,.
\end{eqnarray}
\begin{eqnarray}
\label{f0phi}
B(J/\psi\to f_0(980)\phi)=(3.2\pm 0.9)\cdot 10^{-4}\,.
\end{eqnarray}
\begin{eqnarray}
\label{f2omega}
&& B(J/\psi\to f_2(1270)\omega)=(4.3\pm 0.6)\cdot 10^{-3}\,,
\end{eqnarray}
\begin{eqnarray}
\label{f2'phi}
B(J/\psi\to f_2'(1525)\phi)=(8\pm 4)\cdot 10^{-4}\,,
\end{eqnarray}

The suppression
\begin{eqnarray}
\label{f0omega/f2omega}
&& B(J/\psi\to f_0(980)\omega) /B(J/\psi\to f_2(1270)\omega)= 0.033\pm 0.013
\end{eqnarray}
looks strange in the model under consideration as well as Eq.
(\ref{a0rho/a2rho}) in the model (\ref{a0qq}).

The existence of the $J/\psi\to f_0(980)\phi$-decay of greater intensity than
the $J/\psi\to f_0(980)\omega$-decay ( compare Eq. (\ref{f0omega}) and Eq.
(\ref{f0phi}) ) shuts down the model (\ref{f0qq}) for in the case under
discussion the $J/\psi\to f_0(980)\phi$-decay should be suppressed in
comparison with the $J/\psi\to f_0(980)\omega$-decay by the OZI-rule.

So, Eq. (\ref{f0qq}) is excluded at a level of physical rigor.

Can one consider the $f_0(980)$-meson as the near $s\bar s$-state?

It is impossible without a gluon component. Really, it is anticipated
for the scalar $s\bar s$-state from the lowest P-wave multiplet that
\cite{farrar-94}
\begin{eqnarray}
\label{farrar2s}
&& B(J/\psi\to\gamma f_0(980))\gtrsim B(J/\psi\to\gamma f_2^\prime(1525))/4
\simeq 1.6\cdot 10^{-4}
\end{eqnarray}
opposite Eq. (\ref{f0gluons}), which requires properly that the
$f_0(980)$-meson be the 8-th component of the $SU_f(3)$-oktet
\begin{eqnarray}
\label{oktet}
&& f_0(980)=(u\bar u+d\bar d-2s\bar s)/\sqrt{6}\,.
\end{eqnarray}

This structure gives
\begin{eqnarray}
\label{gf0ggoktet}
&&\Gamma(f_0\to\gamma\gamma)=\frac{3}{25}(1.7 - 5.5)\Gamma
(f_2\to\gamma\gamma)=(0.57 - 1.9)(1\pm 0.14)\,\mbox{keV,}
\end{eqnarray}
that is on the verge of conflict with Eq. (\ref{gf0ggexp}).

Besides, it predicts
\begin{eqnarray}
\label{f0phif0omega}
&&B(J/\psi\to f_0(980)\phi)=(2\lambda\approx 1)\cdot B(J/\psi\to f_0(980)
\omega)\,,
\end{eqnarray}
that also is on the verge of conflict with experiment, compare Eq.
(\ref{f0omega}) with Eq. (\ref{f0phi}).

Eq. (\ref{oktet}) contradicts Eq. (\ref{r}) for the prediction
\begin{eqnarray}
\label{roktet}
R=|g_{f_0K^+K^-}/g_{f_0\pi^+\pi^-}|^2=(\sqrt{\lambda}-2)^2/4\simeq 0.4\,.
\end{eqnarray}

Besides, in this case the mass degeneration $m_{f_0}\simeq m_{a_0}$ is
coincidental, if to treat the $a_0$-meson as the four-quark state, or
contradicts the light hypothesis (\ref{a0qq}).

The introduction of a gluon component, $gg$, in the $f_0(980)$-meson
structure  allows the weak coupling with gluons (\ref{f0gluons}) to be
resolved easy. Really,  by \cite{farrar-94},
\begin{eqnarray}
\label{rtogg}
&& B(R[q\bar q]\to gg)\simeq O(\alpha_s^2)\simeq 0.1 - 0.2\,,\nonumber\\
&& B(R[gg]\to gg)\simeq O(1)\,,
\end{eqnarray}
then the minor ($\sin^2\alpha\leq 0.08$) dopant of the gluonium
\begin{eqnarray}
\label{f0ss}
&& f_0=gg\sin\alpha +\left [\left (1/\sqrt{2}\right )(u\bar u+d\bar d)
\sin\beta + s\bar s\cos\beta\right ]\cos\alpha\,,\nonumber\\
&&\tan\alpha=-O(\alpha_s)\left (\sqrt{2}\sin\beta +\cos\beta\right )\,,
\end{eqnarray}
allows to satisfy Eqs. (\ref{r}), (\ref{f0gluons}) and to get the weak
coupling with photons
\begin{eqnarray}
\label{f0togamagammanearss}
&& \Gamma (f_0(980))\to\gamma\gamma)< 0.22\,\mbox{keV}
\end{eqnarray}
at
\begin{eqnarray}
\label{tgbeta}
&& -0.22>\tan\beta > -0.52\,.
\end{eqnarray}

So, $\cos^2\beta > 0.8$ and the $f_0(980)$-meson is near the $s\bar s$-state,
as in \cite{nils-82}.

It gives
\begin{eqnarray}
\label{f0omega/f0phiss}
&&0.1 < \frac{B(J/\psi\to f_0(980)\omega)}{B(J/\psi\to f_0(980)\phi)}=
\frac{1}{\lambda}\tan^2\beta < 0.54\,.
\end{eqnarray}
As for the experimental value
\begin{eqnarray}
\label{f0omega/f0phiexp}
B(J/\psi\to f_0(980)\omega)/B(J/\psi\to f_0(980)\phi)=0.44\pm 0.2\,,
\end{eqnarray}
it needs refinement.

The scenario, in which with Eq. (\ref{f0ss}) the $a_0(980)$-meson is
the two-quark state (\ref{a0qq}), runs into following difficulties:\\
i) it is impossible to explain the $f_0$ and $a_0$-meson mass degeneration;\\
ii) it is possible to get only
\cite{achasov-89,achasov-97}
\begin{eqnarray}
\label{phigammaf0a0}
&& B(\phi\to\gamma f_0\to\gamma\pi^0\pi^0)\simeq 1.7\cdot 10^{-5}\,,
\nonumber\\
&& B(\phi\to\gamma a_0\to\gamma\pi^0\eta^0)\simeq 10^{-5}\,;
\end{eqnarray}
iii) it is predicted
\begin{eqnarray}
\label{a0gammgammaf0gammagamma}
&&\Gamma(f_0\to\gamma\gamma)<0.13\cdot\Gamma(a_0\to\gamma\gamma)\,,
\end{eqnarray}
that is on the verge of conflict with the experiment, compare Eqs.
(\ref{ga0ggexp}) and (\ref{gf0ggexp});\\
iv) it is also predicted
\begin{eqnarray}
\label{a0rhof0phi}
&& B(J/\psi\to a_0(980)\rho)=(3/\lambda\approx 6)\cdot
B(J/\psi\to f_0(980)\phi)\,,
\end{eqnarray}
that has almost no chance, compare Eqs. (\ref{a0rho}) and (\ref{f0phi}).

Note that the $\lambda$ independent prediction
\begin{eqnarray}
\label{f0phi/f2'phia0rho/a2rho}
&& B(J/\psi\to f_0(980)\phi)/B(J/\psi\to f_2'(1525)\phi)=\nonumber\\
&& = B(J/\psi\to a_0(980)\rho)/B(J/\psi\to a_2(1320)\rho)
\end{eqnarray}
is excluded by the central figure in
\begin{eqnarray}
\label{f0phi/f2'phi}
&& B(J/\psi\to f_0(980)\phi) /B(J/\psi\to f_2'(1525)\phi)= 0.4\pm 0.23\,,
\end{eqnarray}
obtained from Eqs. (\ref{f0phi}) and (\ref{f2'phi}), compare with Eq.
(\ref{a0rho/a2rho}). But, certainly, experimental error is too large.
Even twofold increase in accuracy of measurement of Eq. (\ref{f0phi/f2'phi})
could be crucial in the fate of the scenario under discussion.

The prospects to consider the $f_0(980)$-meson as the near $s\bar s$-state
(\ref{f0ss}) and the $a_0(980)$-meson as the four-quark state (\ref{a0qqqq})
with the coincidental mass degeneration is rather gloomy especially as the
four-quark model with the symbolic structure
\begin{eqnarray}
\label{f0qqqq}
&& f_0=s\bar s(u\bar u+d\bar d)\cos\theta /\sqrt{2}+
u\bar ud\bar d\sin\theta\,,
\end{eqnarray}
built around the MIT-bag \cite{jaffe-77}, reasonably justifies all unusual
features of the $f_0(980)$-meson \cite{achasov-84,achasov-9188}.

Really, the strong coupling with the $K\bar K$-channel is resolved at
$1/16 < \tan^2\theta < 1/2$, see \cite{achasov-84}. There is no problem of
the $a_0$ and $f_0$-meson mass degeneracy at $\tan^2\theta < 1/3$.
The weak coupling with photons was predicted in
\cite{achasov-82}
\begin{eqnarray}
\label{gf0gg4q}
&& \Gamma(f_0(980)\to\gamma\gamma)\sim 0.27\,\mbox{keV.}
\end{eqnarray}
There is also no problem with the suppression (\ref{f0omega/f2omega}).

But, it should be explained how the problem of the weak coupling with gluons
is resolved. Recall that in the MIT-model the $f_0(980)$-meson "consists" of
pairs of colorless and colored pseudoscalar and vector two-quark mesons
\cite{jaffe-77,achasov-82,achasov-84}), including the pair of the flavorless
colored vector two-quark mesons. It is precisely  this pair that converts to
two gluons in the lowest order in $\alpha_s$.

The width of the $f_0(980)$-meson decay in two gluons can be calculated much
as  the width of a four-quark state decay in two photons \cite{achasov-82}.
It gives
\begin{eqnarray}
\label{gf0gg4q}
&& \Gamma (f_0\to gg)=\frac{g_0^2}{16\pi m_{f_0}}0.03
\left (\frac{\alpha_s4\pi}{f^2_{\underline V}}\right )^2(1+
\tan\theta )^2\cos^2\theta \,,
\end{eqnarray}
where $g^2_0/4\pi\approx 10 - 20$ GeV is the OZI-superallowed coupling
constant, 0.03 is the fraction of the pair of the flavorless colored vector
two-quark mesons in the $f_0(980)$-meson wave function, that converts to two
massless gluons, $\alpha_s4\pi/f^2_{\underline V}$ is the probability of
the transition of the flavorless colored vector two-quark meson in  the
massless gluon, ${\underline V}\leftrightarrow g$,
$f^2_{\underline V}/4\pi=f^2_\rho/8\pi\approx 1$ for the space wave functions
of the flavorless colored vector two-quark meson and the $\rho$-meson are
the same. So,
\begin{eqnarray}
\label{gf0gg4q1}
&& \Gamma (f_0\to gg)\approx 15\alpha_s^2(1+\tan\theta )^2\cos^2\theta\ \
\mbox{MeV.}
\end{eqnarray}
At $-1/\sqrt{2}<\tan\theta < -1/4$ one gets the width that is at worst of
order of magnitude less than in the two-quark scalar meson case
\cite{farrar-94} and does not contradict Eq. (\ref{f0gluons}).

If to use only planar diagrams one can get in the four-quark model
\begin{eqnarray}
\label{j/psi4q}
&& B(J/\psi\to a_0^0(980)\rho^0)\approx B(J/\psi\to f_0(980)\omega)\approx
0.5B(J/\psi\to f_0(980)\phi)\,,
\end{eqnarray}
that does not contradict experiment, see Eqs. (\ref{a0rho}), (\ref{f0omega})
and (\ref{f0phi}).

Recall that  almost all four-quark states of the MIT-bag \cite{jaffe-77}
are very broad for their decays into the OZI-superallowed channels. That is
why it is impossible to extract them from the background. Only in the rare
cases on or under the thresholds of the OZI-superallowed decay channels the
"primitive" four-quark states should show up as narrow resonances. This sort
evidence of the MIT-bag, probably, are the $a_0(980)$ and $f_0(980)$-mesons,
as well as the resonance-interference phenomena discovered at the thresholds
of the $\gamma\gamma\to\rho^0\rho^0$ and $\gamma\gamma\to\rho^+\rho^-$
reactions ( see review \cite{achasov-9188} ) and predicted in
\cite{achasov-82}.

A few words on the attractive molecular model, wherein the $a_0(980)$ and
$f_0(980)$-mesons are the bound states of the $K\bar K$-system
\cite{isgur,close}.
This model explains the mass degeneration of the states and their strong
coupling with the $K\bar K$-channel. In the molecular model, as in the
four-quark model, there is no problems with the suppressions
(\ref{a0rho/a2rho}) and (\ref{f0omega/f2omega}). Note that Eq. (\ref{j/psi4q})
is also in the $K\bar K$-molecule model.

But its predictions for two-photon widths \cite{barnes-85}
\begin{eqnarray}
\label{ga0f0ggKK}
&& \Gamma(a_0(K\bar K)\to\gamma\gamma)=\Gamma(f_0(K\bar K)\to\gamma\gamma)
\approx 0.6\,\mbox{keV}
\end{eqnarray}
is on the verge of conflict with the data (\ref{ga0ggexp}) and
(\ref{gf0ggexp}). Besides, the $K\bar K$-molecule widths should be less
bound energy $\epsilon\approx 20$ MeV. The current data \cite{pdg-98},
$\Gamma_{a_0}\simeq 50 - 100\ \ \mbox{MeV}$ and
$\Gamma_{f_0}\simeq 40 - 100\ \ \mbox{MeV}$, contradict this. The
$K\bar K$-molecule model predicts also \cite{achasovmol-97}
\begin{eqnarray}
\label{molphitogammaf0(a0)}
B(\phi\to\gamma f_0\to\gamma\pi\pi)\simeq
B(\phi\to\gamma a_0\to\gamma\pi^0\eta)\simeq 10^{-5}
\end{eqnarray}
that contradict Eqs. (\ref{snd1}) and (\ref{snd2}).

The studies of the $a_0(980)$ and $f_0(980)$-meson production in the
$\pi^-p\to\pi^0\eta n$ \cite{dzierba} and  $\pi^-p\to\pi^0\pi^0 n$
\cite{prokoshkin} reactions over a wide range of the four-momentum transfer
square $0<-t<1\,\mbox{GeV}^2$ show that these states are compact like the
$\rho$ and other two-quark mesons but not extended like the molecules with
the form factors due to the wave functions. It seems that these experiments
leave no chance to the $K\bar K$-molecule model. As for the four-quark
states, they are compact like the two-quark ones.

Lastly, there is a need to answer to the traditional question. Where are the
scalar two-quark states from the lowest P-wave multiplet with the quark
structures (\ref{a0qq}) and (\ref{f0qq})? We believe that there is
no a tragedy with it now! All members of this multiplet are established
\cite{pdg-98}:
\begin{eqnarray}
\label{Pwaves}
&&b_1(1235)\,,\ \ I^G(J^{PC})=1^+(1^{+-})\,,\ \ \Gamma_{b_1(1235)}\simeq 142
\,\mbox{MeV,}\nonumber\\
&&h_1(1170)\,,\ \ I^G(J^{PC})=0^-(1^{+-})\,,\ \ \Gamma_{h_1(1170)}\simeq 360
\,\mbox{MeV,}\nonumber\\
&&a_1(1260)\,,\ \ I^G(J^{PC})=1^-(1^{++})\,,\ \ \Gamma_{a_1(1260)}= 250\,
\mbox{to}\,600\,\mbox{MeV,}\nonumber\\
&&f_1(1285)\,,\ \ I^G(J^{PC})=0^+(1^{++})\,,\ \ \Gamma_{f_1(1285)}\simeq 24
\,\mbox{MeV,}\nonumber\\
&&a_2(1320)\,,\ \ I^G(J^{PC})=1^-(2^{++})\,,\ \ \Gamma_{a_2(1320)}\simeq 107
\,\mbox{MeV,} \nonumber\\
&&f_2(1270)\,,\ \ I^G(J^{PC})=0^+(2^{++})\,,\ \ \Gamma_{f_2(1270)}\simeq 185
\,\mbox{MeV,}\nonumber\\
&&a_0(1450)\,,\ \ I^G(J^{PC})=1^-(0^{++})\,,\ \ \Gamma_{a_0(1450)}\simeq 265
\,\mbox{MeV,} \nonumber\\
&&m_{a_0(1450)}=1300\ \ \mbox{to}\ \ 1500\,\mbox{MeV,}\ \ \ \
\Gamma_{a_0(1450)}=100\ \ \mbox{to}\ \ 300\,\mbox{MeV,}\nonumber\\
&&f_0(1370)\,,\ \ I^G(J^{PC})=0^+(2^{++})\,,\ \ \Gamma_{f_0(1370)}=200\ \
\mbox{to}\ \ 500\,\mbox{MeV,}
\nonumber\\
&&m_{f_0(1370)}=1200\ \ \mbox{to}\ \ 1500\,\mbox{MeV.}
\end{eqnarray}
Certainly, one cannot consider the scalar members of Eq. (\ref{Pwaves}) as
much-established and it needs else the refinement of their masses and widths.
Nevertheless, from  Eq. (\ref{Pwaves}) it will be obvious that forces, responsible for
splitting of masses in the P-wave multiplet, are either small or compensate
each other.  That is why we rightfully expect the existence of the
$a_0(\approx 1300)$ and $f_0(\approx 1300)$-states and it seems by far that
the $a_0(980)$ and $f_0(980)$-mesons are foreigners in the company
(\ref{Pwaves}).

The statement of the OPAL Collaboration \cite{lafferty-98} on a consistency
of the $f_0(980)$ inclusive production in hadronic $Z^0$ decay with the
hypothesis (\ref{f0qq}) is not conclusive because any calculations of the
inclusive production of four-quark states are absent. Nevertheless, from
general point of view one can expect as copious multiquark state inclusive
production as two-quark one because in both cases the primary production is
the multiproduction of the vacuum $q\bar q$ pairs.

\section{The theoretical grounds for the four-quark model}

A few words on the theoretical grounds for the four-quark nature of the scalar
$f_0(980)$ and $a_0(980)$ mesons. It was shown  in the context of the MIT-bag
\cite{jaffe-77} that the low-lying scalar four-quark nonet as bound state of
diquarks ($T_a=\varepsilon_{abc}\bar q^b\bar q^c$ and $\bar T^a=
\varepsilon^{abc}q_bq_c$, note that similar diquarks binding up with quarks
to form the baryon octet) arises from the strong binding energy in such a
configuration due to a hyperfine interaction Hamiltonian of the form
\begin{eqnarray}
&&H_{hf}=-\Delta\sum_{i<j}\vec s_i\cdot\vec s_j\vec F_i\cdot\vec F_j\,,
\quad\vec s=\frac{\vec\sigma}{2}\,,\; \vec F=\frac{\vec\lambda}{2}\,,
\nonumber
\end{eqnarray}
that obtained from one gluon exchange in QCD.

The result is \cite{jaffe-77}
\begin{eqnarray}
&&\sigma=C^0=ud\bar u\bar d\,; \quad\kappa^+=C_{K^+}=ud\bar d\bar s\,,
\quad\kappa^0=C_{K^0}=ud\bar u\bar s\,, \nonumber\\
&&\kappa^-=C_{K^-}=ds\bar u\bar d\,,\quad\bar\kappa^0=C_{\bar K^0}=
us\bar u\bar d\,;\quad f_0=C^s=
\frac{(us\bar u\bar s+ds\bar d\bar s)}{\sqrt{2}}\,;
\nonumber\\
&& a_0^+=C^s_{\pi^+}=us\bar d\bar s\,,\quad a_0^0=C^s_{\pi^0}=
\frac{(us\bar u\bar s-ds\bar d\bar s)}{\sqrt{2}}\,,\quad a_0^-=C^s_{\pi^-}=
ds\bar u\bar s\,,\nonumber\\
&& m(\sigma)=650\mbox{MeV}\,,\quad m(\kappa)=900\mbox{MeV}\,,\quad
m(f_0)=m(a_0)=1100\mbox{MeV}\,.
\nonumber
\end{eqnarray}

Certainly, the MIT-bag model is rather rough, so that one can consider its
prediction only as a guide. As the $\sigma$ and $\kappa$ mesons lie
considerably above their the OZI-superallowed channels their widths have of
the 1 GeV order. That is why the information about them can be got only
in a model dependent way.

In the last few years the true renaissance has been going in treatments of
$\pi\pi$ and $\pi K$ scattering with help of phenomenological linear $\sigma$
models, see, for example,
\cite{achasov-94,schechter-95,tornqvist-95,scadron-95,ishida-96}.
It has been argued on occasion that the corresponding $\sigma$ mesons are
quark-antiquark states. But in fact at the Lagrangian level there is no
difference in the formulation of the two-quark and four-quark cases
\cite{schechter-99}.

\section{Conclusion}

So, there are many reasons to consider the $a_0(980)$ and $f_0(980)$ mesons
as the four-quark states. Nevertheless, in summary one emphasizes once again
that the further study of the $\phi\to\gamma f_0(980)$, $\gamma a_0(980)$,
$J/\psi\to a_0(980)\rho$, $f_0(980)\omega$, $f_0(980)\phi$, $a_2(1320)\rho$,
$f_2(1270)\omega$, $f_2'(1525)\phi $, $a_0(980)\to\gamma\gamma$ and
$f_0(980)\to\gamma\gamma$ decays will enable one to solve the question on
the $a_0(980)$ and $f_0(980)$-mesons nature, at any case to close the above
scenarios.

I thank V.V. Gubin and G.N. Shestakov for useful discussions.

 The present work was supported in part by the grant INTAS-94-3986.

\end{document}